\documentclass[]{elsart}

\usepackage[dvips]{graphicx}
\usepackage{amsmath}
\usepackage{amssymb}
\usepackage{psfrag}

\newcommand{\sR}{\mathcal{R}}
\newcommand{\sT}{\mathcal{T}}

\newcommand{\averfull}[1]{\langle #1 \rangle}
\newcommand{\heta}{\hat{\eta}}
\newcommand{\hHU}{\widehat{HU}}
\newcommand{\hHV}{\widehat{HV}}
\newcommand{\hHW}{\widehat{HW}}
\newcommand{\hu}{\hat{u}}
\newcommand{\hv}{\hat{v}}
\newcommand{\hw}{\hat{w}}

\begin{document}
\begin{frontmatter}
\title{Direct Numerical Simulation \\ of turbulent Taylor--Couette flow}
  \author[aut1]{Davide Pirr\`o, Maurizio Quadrio\corauthref{corr1}}
  \ead{davide.pirro@polimi.it}
  \ead{maurizio.quadrio@polimi.it}

  \address[aut1]{Dipartimento di Ingegneria Aerospaziale Politecnico
    di Milano - Italy}

  \corauth[corr1]{Corresponding author. Address: Dipartimento di
    Ingegneria Aerospaziale del Politecnico di Milano - via La Masa 34
    - 20156 Milano - Italy}

\begin{abstract}
The direct numerical simulation (DNS) of the Taylor--Couette flow in the fully turbulent regime is described. The numerical method extends the work by Quadrio \& Luchini (Eur. J. Mech. B / Fluids, {\bf 21}, 413--427, 2002), and is based on a parallel computer code which uses mixed spatial discretization (spectral schemes in the homogeneous directions, and fourth-order, compact explicit finite-difference schemes in the radial direction). A DNS is carried out to simulate for the first time the turbulent Taylor--Couette flow in the turbulent regime. Statistical quantities are computed to complement the existing experimental information, with a view to compare it to planar, pressure-driven turbulent flow at the same value of the Reynolds number. The main source for differences in flow statistics between plane and curved-wall flows is attributed to the presence of large-scale rotating structures generated by curvature effects.
\end{abstract}
\end{frontmatter}

\section{Introduction}
The flow in the gap between coaxial rotating cylinders, known as the Taylor--Couette flow (TC flow hereinafter), is among the most investigated problems in fluid mechanics (see \cite{tagg-1994} for an overview), owing to its engineering applications \cite{haim-pismen-1994}, as well as to its relevance as prototypical flow in the study of transition to turbulence and of fully-developed turbulent flows over streamwise-curved surfaces.

Many experimental and numerical papers have dealt with the instabilities developed by the TC flow when the value of the Reynolds number $Re$ is increased such that the laminar regime is taken over by a sequence of bifurcations, eventually leading to chaotic behavior. Less is known about the fully turbulent regime, since a smaller number of laboratory experiments and essentially no direct numerical simulation (DNS) studies are available. The lack of DNS investigations can be ascribed -- at least in part -- to the difficulties of implementing an efficient method for the numerical solution of the incompressible Navier--Stokes equations in cylindrical geometries, and at the same time to the computational demand of such simulations in turbulent regime. Hence a satisfactory quantitative description of the turbulent TC flow, leveraging the full space-time information potentially available from a DNS, is as yet missing.

In the first part of this paper a numerical method for the DNS of the Navier--Stokes in cylindrical coordinates is presented, which is designed for computing the Taylor--Couette flow in the turbulent regime. The method relies on the strategy developed by Quadrio \& Luchini \cite{quadrio-luchini-2002} for the DNS of a turbulent flow in an annular pipe: the main enhancements introduced here are an improved accuracy of the spatial discretization, and the ability of the code to exploit parallel computing on commodity hardware.

The newly developed numerical method is then used to carry out what we believe is the first DNS of the turbulent Taylor--Couette flow. We shall focus on a so-called small-gap geometry, identical to that considered by Andereck, Liu \& Swinney \cite{andereck-liu-swinney-1986}, and set the Reynolds number to a relatively high value. Analysis of the results will focus on the assessment of low-order flow statistics, and on some theoretical issues concerning turbulence in presence of wall curvature. As  described at length in the review paper \cite{patel-sotiropoulos-1997} by Patel \& Sotiropoulos, curvature considerably impacts the modeling of turbulent flows. Its effects thus carry a broad interest, and also affect experimental practices. For example, when measuring the skin friction in a turbulent flow over a curved wall, one often resorts to methods (like the Clauser plot) which imply the validity of the law of the wall, and which often require the numerical values of its parameters (i.e. the slope of the logarithmic part of the profile, given by the inverse of the von K\'arm\'an constant, and its intercept) to be known in advance. Indeed, neither these values possess to date an agreed-upon value, nor is their dependence on the degree of curvature known. A further example occurs in the derivation of the friction law for the TC flow, i.e. a formula relating the friction coefficient to the value of the Reynolds number: such derivation, actively discussed in literature \cite{lathrop-fineberg-swinney-1992,panton-1992,lewis-swinney-1999,eckhardt-grossmann-lohse-2000}, is often based on assumptions (logarithmic form vs. power law) about the shape of the mean velocity profile. A DNS, with its ability to evaluate the skin friction directly, is a powerful tool to complement experimental data in this research areas. Of course the main limitation of DNS lies in the limited values of the Reynolds number typically affordable in the numerical simulations. This is the reason why in this paper the numerical method is developed with great emphasis on computational efficiency.

A DNS at a relatively high value of $Re$ will also offer the opportunity to observe to what extent a planar, pressure-driven and fully-developed turbulent flow (like a channel flow) presents analogies to the TC flow, in the spirit of what has been already observed by Lee \& Kim \cite{lee-kim-1991} for the planar TC flow. The present simulation reaches an unprecedently high value of $Re$, so that the comparison is made at the value $Re_\tau \approx 180$ (based on the friction velocity and half the gap width), which is considered the minimum bound for the near-wall viscous turbulent cycle to be fully active. A significant physical process which acts in the TC flow and is absent in plane geometries is the additional mass and momentum transfer exerted by large-scale toroidal rolls. The laminar solution is known to be unstable above a well-defined (and curvature-dependent) critical value $Re_c$ of the Reynolds number, above which such structures, known as Taylor vortices, are quickly generated \cite{taylor-1936}. As $Re$ increases further, the Taylor vortices undergo a series of transformations, after which they eventually reappear in the turbulent regime \cite{koschmieder-1979}. In his illuminated paper \cite{townsend-1984}, Townsend surmised that turbulence in the TC flow is basically of two different kinds, with one contribution from the wall shear and the other from the large-scale structures. DNS is the perfect tool to analyze the full flow field and to describe these large-scale structures, that are absent both in the plane channel and in the plane Couette flows.

The outline of the paper is as follows. In \S\ref{sec:method} the geometry of the problem is introduced, and the main elements of its numerical simulation are recalled in \S\ref{sec:equations} (additional information is deferred to an Appendix). A validation of the computer code and a critical comparison of its results to available data will be given in \S\ref{sec:validation}, together with a quantitative assessment of the computational performance. The numerical simulation of the turbulent TC flow will be described in \S\ref{sec:turbulent-dns}, with regard to discretization parameters and computational procedures. In \S\ref{sec:results} results will be presented in terms of averaged and instantaneous flow properties. Lastly, \S\ref{sec:conclusions} will contain a conclusive summary.

\section{Problem definition and numerical method}
\label{sec:method}

\begin{figure}
\psfrag{Wi}{$W_i$}
\psfrag{Lx}{$L_x$}
\psfrag{Lt}{$L_\theta$}
\psfrag{x,u}{$x,u$}
\psfrag{r,v}{$r,v$}
\psfrag{theta,w}{$\theta,w$}
\psfrag{Ri}{$\sR_i$}
\psfrag{Ro}{$\sR_o$}
\psfrag{2d}{$2h$}
\centering
\includegraphics[width=\columnwidth]{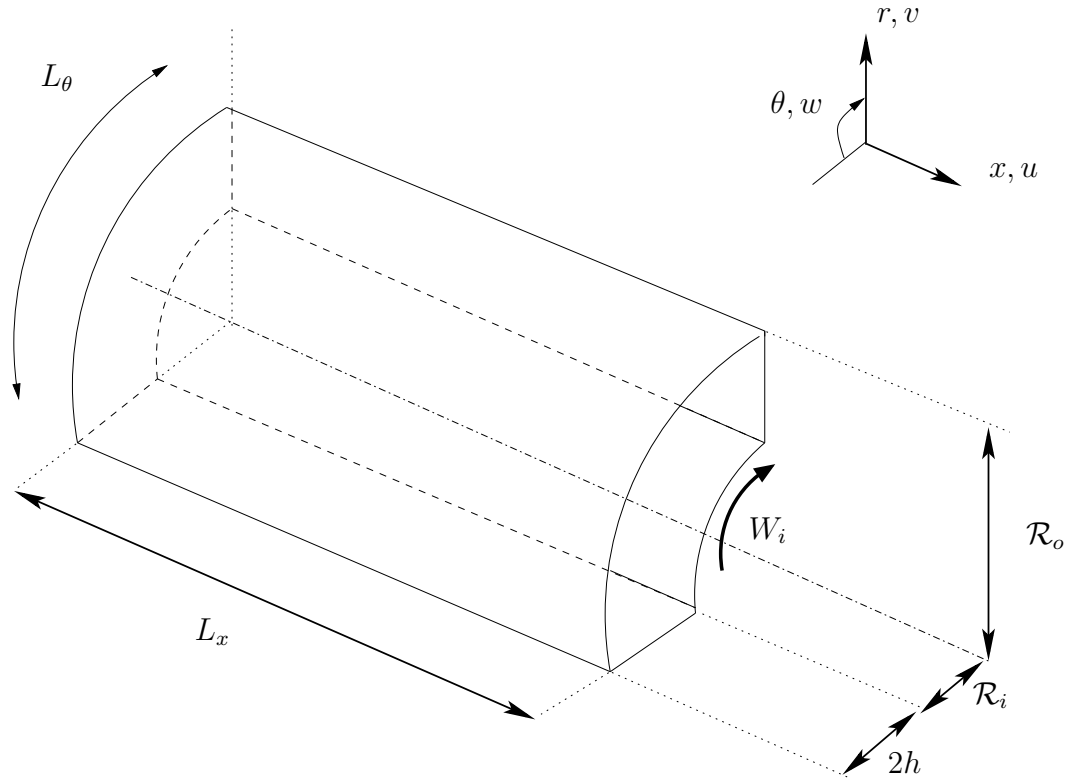}
\caption{Geometry of the Taylor--Couette flow system. The inner cylinder rotates at speed $W_i$ and angular velocity $\Omega_i$, the outer cylinder is at rest.}
\label{fig:geometry}
\end{figure}
We consider the classic Taylor--Couette apparatus, made by a moving inner cylinder with radius $\sR_i$ rotating at an angular velocity $\Omega_i$, and a concentric, outer fixed cylinder with radius $\sR_o$. In figure \ref{fig:geometry} the geometry employed in the present work is sketched: the radial coordinate is $r$, the axial and azimuthal coordinates being $x$ and $\theta$. The corresponding velocity components are $v$ (wall-normal), $u$ (spanwise) and $w$ (streamwise). The amount of curvature is expressed by the geometric parameter $\zeta=\sR_i/\sR_o$. The Reynolds number $Re$ is defined with the inner cylinder rotating speed $W_i=\Omega_i \sR_i$, the gap width $\sR_o - \sR_i = 2h$ between the two cylinders and the kinematic viscosity $\nu$ of the fluid. For a given value of $\zeta$, when gravity and other external forces are neglected, $Re$ becomes the sole relevant fluid dynamic parameter. Unless otherwise stated, the results presented in the following are made non-dimensional by using $W_i$ as reference velocity and $h$ as reference length. When useful, the wall distance $y$ is also used, conveniently defined as $y=r-\sR_i$ or $y=\sR_o -r$. Quantities made non-dimensional in wall units are indicated with a $+$ superscript and are computed with friction velocity $u_\tau$ (see \S\ref{sec:results} for details concerning its definition) and fluid viscosity $\nu$.

\subsection{Governing equations and discretization}
\label{sec:equations}

The Taylor--Couette system is of course most easily described in cylindrical coordinates. On the other hand, a widely used and extremely efficient formulation of the Navier--Stokes equations exists in cartesian coordinates, and is well suited to the DNS of turbulent flows with two homogeneous directions. This formulation was introduced by Kim, Moin \& Moser \cite{kim-moin-moser-1987}, and is based on rewriting the equations of motion in terms of two scalar equations for the wall-normal components of the velocity and vorticity vector. The computational efficiency is maximized when flow variables are expanded in Fourier series in the two homogeneous directions (the wall-normal discretization is irrelevant).

A few studies have described the extension of this formulation to a cylindrical geometry, to enjoy the same advantages: pressure removed from the equations, and optimal computational load. We build for the present work upon the contribution by Quadrio \& Luchini \cite{quadrio-luchini-2002}, who wrote evolutive equations for the radial components of velocity and vorticity in cylindrical coordinates, and used finite differences to compute radial derivatives with schemes of second-order formal accuracy. The time integration was carried out in \cite{quadrio-luchini-2002} with an explicit third-order Runge--Kutta method for the nonlinear terms and curvature-related viscous terms, and a second-order Crank--Nicholson scheme for the remaining viscous terms.

The present work employes a numerical method which is improved in at least two respects compared to Ref. \cite{quadrio-luchini-2002}: the second-order accuracy of the finite-difference schemes was not considered satisfactory, and -- most importantly -- distributed-memory parallel computing capabilities are required. As shown by Luchini and Quadrio in their recent paper \cite{luchini-quadrio-2006}, these two issues can be dealt with in an unified approach. The formal accuracy of the radial derivatives -- and, most importantly, their spectral resolution -- is improved by resorting to high-accuracy compact difference schemes. The schemes are moreover made explicit, by following the procedure illustrated by Thomas \cite{thomas-1953} which leverages the absence of the third derivative in the equations of motions, to reduce their computational cost. At the same time, finite differences, which enjoy the property of being local operators in physical space, are key to obtain a good parallel performance on commodity networking hardware.

Reproducing the approach of Ref. \cite{luchini-quadrio-2006} in the cylindrical case, however, is not trivial. As shown in the Appendix, the governing equations require further manipulation in order to arrive at a form where explicit compact schemes can be applied. A computational stencil made by 5 arbitrarily spaced (and smoothly stretched) grid points is used to obtain a formal accuracy of order at least 4, and the coefficients are computed, according to an approach common in the theory of Pad\'e approximants, by matching the discrete derivative of a function known in analytic form (e.g. a polynomial) to its analytical derivative. We refer the reader to Ref. \cite{luchini-quadrio-2006} for a general illustration of the method in the cartesian case, and to the Appendix for a short discussion of the peculiarities of the cylindrical case. Full description of the method is given by Ref. \cite{pirro-2005}.

\section{Validation}
\label{sec:validation}

The present computer code is carefully validated against both numerical and experimental data available for the TC flow through preliminary calculations at non-turbulent values of $Re$.

The correctness and accuracy of finite-difference radial operators is preliminarily checked by computing the exact laminar solution. In the laminar regime, the flow is described by an exact solution of the Navier--Stokes equations \cite{batchelor-1967}:
\begin{equation}
\label{eq:laminar-solution}
w(r)=\frac{W_i}{1-\zeta^2} \left[ \frac{\sR_i}{r} - \zeta^2 \frac{r}{\sR_i} \right].
\end{equation}

Due to the term $\sim r^{-1}$, this solution cannot be represented exactly by the polynomial interpolation implied by a FD method. A discretization error is thus always present, which is bound to decrease as the step size raised to the fourth power.
\begin{figure}
\includegraphics[width=\columnwidth]{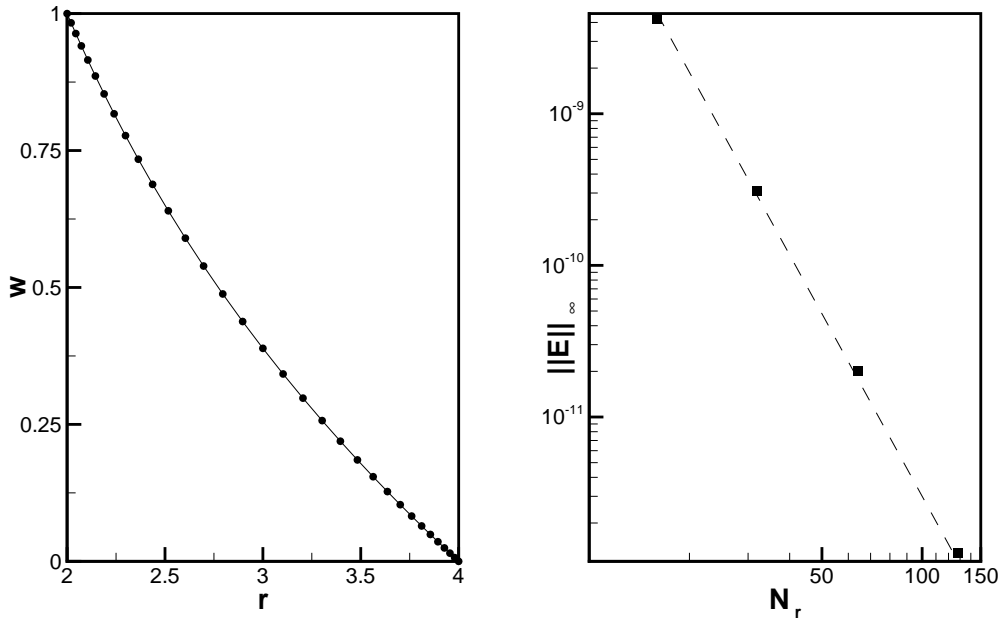}
\caption{Left: computed laminar velocity profile (symbols) and laminar analytical solution (\ref{eq:laminar-solution}) (continuous line), for $\zeta=0.5$ and $N_r=33$ points in the radial direction. Right: difference $||E||_\infty$ between computed and analytical solutions, as a function of $N_r$. The straight line shows the expected decrease proportional to $N_r^{-4}$.}
\label{fig:validation}
\end{figure}
The laminar velocity profile computed for $\zeta=0.5$ is shown in Fig. \ref{fig:validation} together with the analytical solution. The difference between computed and analytical curves is indeed observed to decrease as requested by the formal fourth-order accuracy of the numerical method.

The entire code is then tested both in large- and small-gap geometries for the first insurgence of Taylor vortices, to verify that the evolution of small-amplitude disturbances is well represented. The curvature-dependent critical value $Re_c$ of the Reynolds number, above which instability amplifies flow perturbations to develop Taylor vortices, is computed with the present code and compared to available data. For a given $\zeta$, the code is run at various values of $Re$, starting from an initial condition made by the laminar solution with superimposed small-amplitude random disturbances. The discretization parameters are reported in table \ref{tab:parameters-validation} (``straight-vortex'' case). The temporal evolution of the kinetic energy is monitored, to verify whether the initial energy decreases or gets amplified depending on $Re$. For $\zeta=0.5$ we find $Re_c$ to be confined within the range $68.2 < Re_c < 68.4$. At the same curvature, previous numerical simulations by Fasel \& Booz \cite{fasel-booz-1984}, based on an axisymmetrical finite-difference method, determined $Re_c = 68.2$, whereas experimental measurements \cite{schultz-pfister-2000} indicated $Re_c = 68.4$. For the small-gap case, at $\zeta=0.95$ we compute $184 < Re_c < 186$, which agrees with the value of $185$ determined by Moser et al. \cite{moser-moin-leonard-1983}.

In the same flow regime, the torque needed to maintain the rotation of the inner cylinder is computed as an indicator of global spatial accuracy, and compared with previous works. The torque is defined as:
\begin{equation}
\label{eq:torque}
\sT = \mu 2 \pi \sR_i^2 L \left( \left| \frac{\partial w}{\partial r} \right|_{r=\sR_i} - \frac{w}{\sR_i} \right) .
\end{equation}
where $L$ is the axial length of the cylinders.

Table \ref{tab:torque} show how values of $\sT$ computed by the present code at $\zeta=0.5$ compare very well (within small fractions of a percent) to literature data. 

A further increase in $Re$ brings us into the non-linear regime, where the straight Taylor vortices undergo a wavy azimuthal deformation and are named "wavy vortices" after the paper \cite{coles-1965} by Coles. The prediction of quantitative properties of the wavy Taylor vortices, like their azimuthal wavenumber and their phase speed, allows a further confirmation of the present code. A simulation with the discretization parameters reported in table \ref{tab:parameters-validation} (``wavy-vortex'' case) has been carried out at $\zeta=0.5$ and $Re=250$. In this simulation only, to ease the comparison with the paper \cite{snyder-1969} by Snider, the outer cylinder possesses a small amount of counter-rotation, such that $Re$ based on the outer cylinder's rotation speed is $Re=-55$. The simulation, described in detail in \cite{pirro-2005}, shows a pair of wavy vortices, with azimuthal wavenumber $m=2$, and a characteristic rotation period $T = 117 h / W_i$. This translates into a non-dimensional rotation speed $s = 2 \pi / m T \Omega_i = 0.054$, which compares very well to the results of Ref. \cite{snyder-1969}.

\subsection{Computational performance}
\label{sec:parallel-performance}

Here the computational performance of the code is documented for the problem size and on the computing system used for the turbulent simulation described later in \S\ref{sec:turbulent-dns}.

We employ a spatial resolution of $N_\theta = 512$ and $N_z = 170$ Fourier modes in the azimuthal and axial directions respectively, while the number of points in the radial direction is set to $N_r = 129$. Parallel simulations are carried out on a dedicated computing system, called a Personal Supercomputer, which is based on commodity personal computers. The special architecture of this system, that can be used with either the cartesian or the cylindrical code, has been described in Ref. \cite{luchini-quadrio-2006}. The system is made by 10 SMP nodes, each equipped with 2 Intel Xeon 2.66 GHz CPU, and at least 1GB of 266 MHz SDRAM. The nodes carry two Gigabit Ethernet adapters each, and are connected in a ring topology.

On one Xeon CPU the code requires 680 MB of memory, and takes 182 seconds to compute a full time step for a 3-substeps Runge-Kutta temporal scheme. The amount of memory includes storage space for the radial coefficients discussed in the Appendix. Should RAM size become an issue, these coefficients can can be computed on-the-fly, thus saving 13\% of RAM at the expense of a comparable increase in computing time. The CPU overhead of the cylindrical code compared to its cartesian counterpart is about 40\% overall.

Table \ref{tab:speedup} documents how the computation is speed up by employing an increasing number $N$ of computing nodes. The ratio between the wall-clock time on one CPU and the wall clock time with $N$ distributed-memory CPUs, i.e. the parallel speedup $S$, is not far from the linear one: the PLS parallel strategy in the present context thus exploits fully our low-cost computing system. The maximum measured speedup with $N=10$ is about 9. The SMP capabilities of the nodes are then exploited on top of the distributed-memory parallel strategy. This allows a further 1.6 speedup factor from the use of 2 CPUs. The SMP speedup is additive to the distributed-memory speedup, and shows essentially no decrease with increasing the number of computing nodes. The entire wall clock time for the simulation described in \S\ref{sec:turbulent-dns} can thus be decreased from about 6 months to about 11 days when the entire system is used.

\section{The turbulent Taylor--Couette flow simulations}
\label{sec:turbulent-dns}

The geometry considered in the present analysis is the small-gap geometry described by Andereck, Liu \& Swinney \cite{andereck-liu-swinney-1986} with $\zeta=0.882$, which gives $\sR_i = 15 h$ and $\sR_o = 17 h$. The value of the Reynolds number is set at $Re=10500$, to obtain a value of the friction Reynolds number similar to $Re_\tau=180$, which is typically considered the lowest value at which a turbulent channel flow presents a well-developed inner layer \cite{kim-moin-moser-1987,moser-kim-mansour-1999}. To our knowledge, this is the highest value of $Re$ ever reached in a DNS of Couette flow, and of course represents a big computational challenge: for comparison, Bech et al. \cite{bech-etal-1995} computed a fully turbulent plane Couette flow, which is the limit for $\zeta \rightarrow 1$ of the present TC flow, at $Re_\tau \approx 80$ only. According to Hamilton, Kim \& Waleffe \cite{hamilton-kim-waleffe-1995}, the plane Couette flow is turbulent above $Re_\tau = 30$.

The size of the computational domain must be chosen by keeping in mind that periodic boundary conditions are used in the two homogeneous directions: a periodic box imposes an artificial large-wavelength cutoff to the structures that can be represented in the numerical simulation. It can be recalled however that periodicity in the streamwise direction is a physically sound condition for the TC flow, unlike planar channel and Couette flows. The azimuthal extension of the box is chosen so that $L_{\theta} = 10 \pi h$ when measured at the centerline of the gap. The turbulent TC flow presents a dominant periodic pattern in the axial direction too: toroidal structures present an axial wavelength of $5h$, and experimental results \cite{koschmieder-1979} indicate that this wavelength is rather insensitive to the value of $Re$. The simulations are thus carried out for $L_z = 5 h$. It can be mentioned moreover that the implied indefinite mirroring of periodic boxes makes the present simulation totally free from the end-effects that are unavoidable in laboratory experiments.

To obtain the resolution of all the significant spatial scales, $N_\theta = 512$ and $N_z = 170$ Fourier modes are used in the azimuthal and axial directions respectively, and for the radial direction $N_r = 129$. The spatial resolution in wall units matches or exceeds the typical resolution employed in channel-flow DNS \cite{moser-kim-mansour-1999}. The global number of degrees of freedom amounts to $\approx 2.2 \cdot 10^7$. The parameters related to the spatial discretization are reported in table \ref{tab:parameters-turbulent}. The resolution of the relevant temporal scales dictates the time step size. We use $\Delta t = 0.012 \delta / W_i$, which is smaller than the stability limit of the employed Runge--Kutta scheme. In viscous units, this corresponds to $\Delta t^+ \approx 0.08$. A null mean pressure gradient is imposed in both the homogeneous directions.

The adequacy of the present size for the computational domain can be first evaluated in terms of inner (viscous) units, i.e. by making lengths non-dimensional with the friction velocity $u_\tau$ (see later \S\ref{sec:results} for its definition) and the fluid viscosity. With the friction velocity computed at the inner wall $L_z^+ = 946$, whereas at the outer wall $L_z^+=840$. The streamwise length is between $L_\theta^+ = 5576$ at the inner wall and $L_\theta^+=5600$ at the outer wall. A comparison with the domain sizes usually employed for DNS in the turbulent plane channel flow show that the domain size is certainly adequate for representing the inner turbulent layer. The streamwise length $L_\theta^+$ is 2.5 times the streamwise length typically employed \cite{kim-moin-moser-1987} in the DNS of a plane turbulent channel flow at a comparable value of friction Reynolds number; the spanwise width is large enough to represent well the underlying turbulent flow, being larger than that used in Ref. \cite{moser-kim-mansour-1999}.

It remains to be determined whether representing a single pair of periodic structures is actually enough to obtain domain-independent low-order statistics. Given the emphasis of the present work towards high possible $Re$, we decided to use the available computational resources to reach the highest $Re$ while representing a single pair of TV, and to check a posteriori that low-order statistics are not significantly affected by the axial domain size. The check consisted in running an additional simulation for a domain size with doubled $L_z$, and an accordingly doubled $N_z$. At the same time, the averaging time is halved to maintain the computational load approximately constant. In short, the result of this check, that will be further discussed in the following, is that representing a single pair of structures is adequate for measuring low-order statistics. 

\subsection{Computational procedures}
\label{sec:procedures}

The initial condition for the simulations is based on the laminar solution (\ref{eq:laminar-solution}); divergence-free velocity disturbances with amplitude $\mathcal{O}(10^{-4})$ and random phase are added to the whole set of Fourier modes. Different initial conditions have been considered in (less resolved) preliminary simulations, by either changing the amplitude of the disturbances, or by applying disturbances to a partial set of modes, or by starting the run from a null mean velocity profile instead of the profile given by Eq. (\ref{eq:laminar-solution}). In terms of the long-time mean turbulent friction and others higher-order statistics, no difference has been noticed among these cases after the turbulent regime sets up, even though the time required to reach the statistically steady-state, as well as the behaviour of the flow during the initial transient, have been observed to depend on the details of the initial conditions.

\begin{figure}
\centering
\includegraphics[width=\columnwidth]{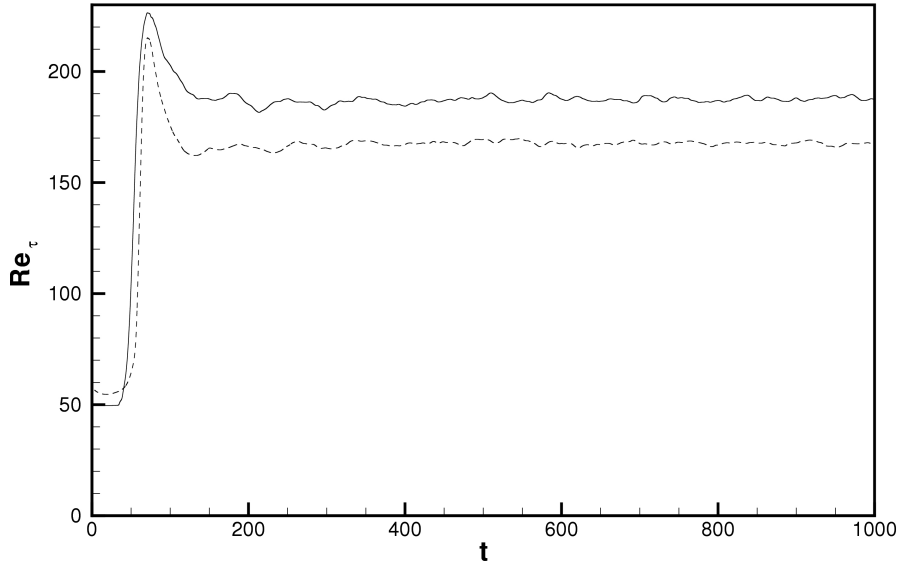}
\caption{Behaviour of $Re_\tau$ at the inner wall (continuous line) and outer wall (dashed line) as a function of time after the startup of the simulation.}
\label{fig:Rtau}
\end{figure}
The simulation is started from the above-described initial condition and let run for $1000 \delta/W_i$. A flow field is stored on disk every 10 time units for later postprocessing. The size of a single flow field is 185 MBytes.

The time required for the simulation to settle to a statistically steady-state requires an initial time interval: we have estimated it based on the time history of various quantities, including the derivative of the space-mean velocity longitudinal velocity profile friction computed at the walls, which is related to $Re_\tau$. Figure \ref{fig:Rtau} shows how the length of this transient is about 150 time units. In computing statistical quantities, flow fields corresponding to the initial 300 time units are rejected.

\section{Results}
\label{sec:results}

The first global flow feature that we discuss  is the mean friction. The shear-stress $\tau_w$ is computed for both the inner and the outer walls. Its mean value $\averfull{\tau_w}$, averaged over time and over the homogeneous directions, allows us to compute a friction velocity $u_\tau=\sqrt{\averfull{\tau_w} / \rho}$ local to each wall. From $u_\tau$ a local Reynolds number $Re_\tau=u_\tau \delta / \nu$ can be obtained. The present results yield $Re_\tau=189.3$ for the inner wall and $Re_\tau=167.7$ for the outer wall. As a further confirmation of the adequacy of spatial resolution and statistical averaging, we observe that the ratio between the values of $Re_\tau$ at the two walls equals $\zeta$ within 0.18\%. The check with larger $L_z$ yielded values of $Re_\tau$ which differ from the above ones by less than 1\%, i.e. less by the error implied by the finite averaging time.

\begin{figure}
\centering
\includegraphics[width=\columnwidth]{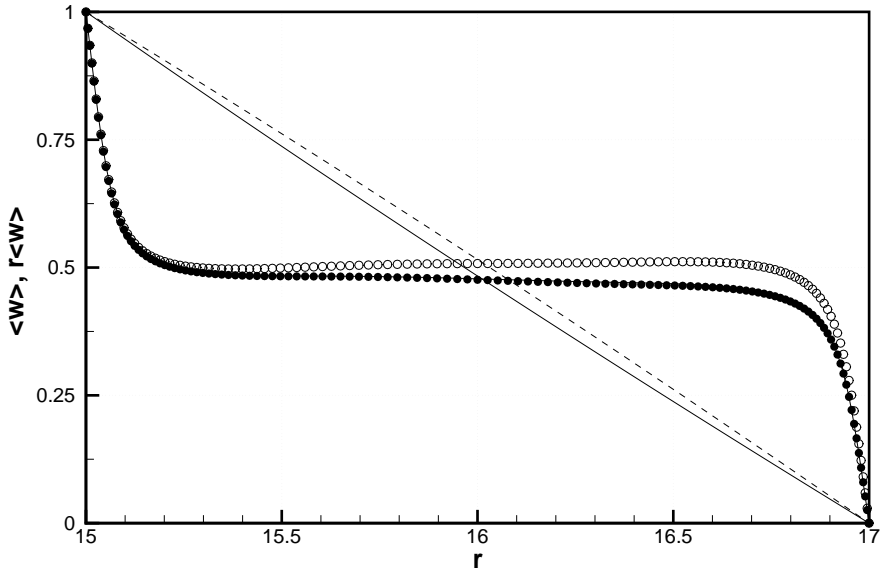}
\caption{Radial distribution of the mean azimuthal velocity $\averfull{w}$ (closed symbols, continuous line) and of the angular momentum $r \averfull{w}$ (open symbols, dashed line). Lines correspond to the analytical laminar solution. Note that angular momentum is made non-dimensional by $W_i \sR_i$. The virtually indistinguishable line drawn along the closed circles is the $\averfull{w}$ profile for the simulation with $L_z=10h$.}
\label{fig:uav} 
\end{figure}
The radial profile $\averfull{w}$ of the mean azimuthal component of the velocity vector is shown in figure \ref{fig:uav} and compared to the laminar solution (\ref{eq:laminar-solution}). While the laminar profile at this curvature level is little different from a straight line, the turbulent profile presents, in analogy to the plane turbulent Couette flow, two distinct regions: near the walls a shear-driven boundary layer, and in the central part of the gap a region where the velocity decreases very slowly with $r$. In particular, the large central region presents an almost constant or very slowly-increasing angular momentum $r \averfull{w}$; its value of nearly $0.5 \sR_i W_i$ agrees with the measurements by Taylor \cite{taylor-1936} and Smith \& Townsend \cite{smith-townsend-1982}. This property of the flow, observed here for the first time in a numerical simulation, qualitatively supports the hypothesis of a core region of constant angular momentum flow put forward by Townsend \cite{townsend-1976}. The figure confirms moreover that the mean profile is essentially unaffected by the number of represented TV pairs: the profile computed for the simulation with $L_z = 10 h$ is hardly distinguishable from the one with $L_z=5h$.

\begin{figure}
\includegraphics[width=\columnwidth]{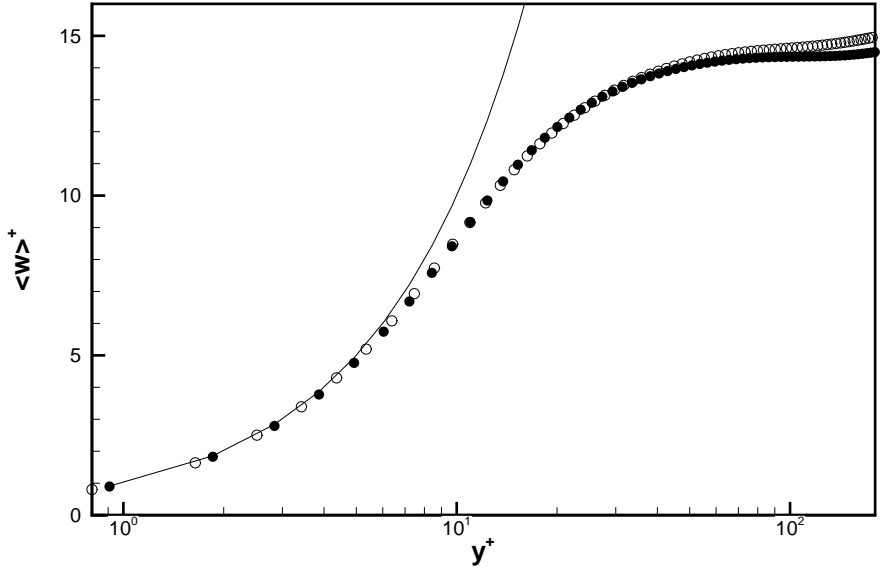}
\centering
\caption{Mean azimuthal velocity profile $\averfull{w}^+$, in local wall units, over the inner wall (closed symbols) and the outer wall (open symbols). The continuous line is the linear profile $\averfull{w}^+ = y^+$.} \label{fig:lotw}
\end{figure}
In figure \ref{fig:lotw} the azimuthal mean velocity profile is plotted in semi-logarithmic scale and in local wall units against the wall distance $y^+$. Similarly to pressure-driven flows, a viscous sublayer can be observed to develop over both walls for $y^+ \lesssim 5$, where the profile follows the linear law $\averfull{w}^+ = y^+$ widely accepted for pressure-driven flows. The profiles over both walls are fairly similar up to $y^+ \approx 40$ (buffer layer), but in the central part of the gap the outer profile becomes slightly higher than the inner one: this indicates that the friction velocity (local to each wall) alone is not the correct scaling velocity in this region of the flow.

\begin{figure}
\centering
\includegraphics[width=\columnwidth]{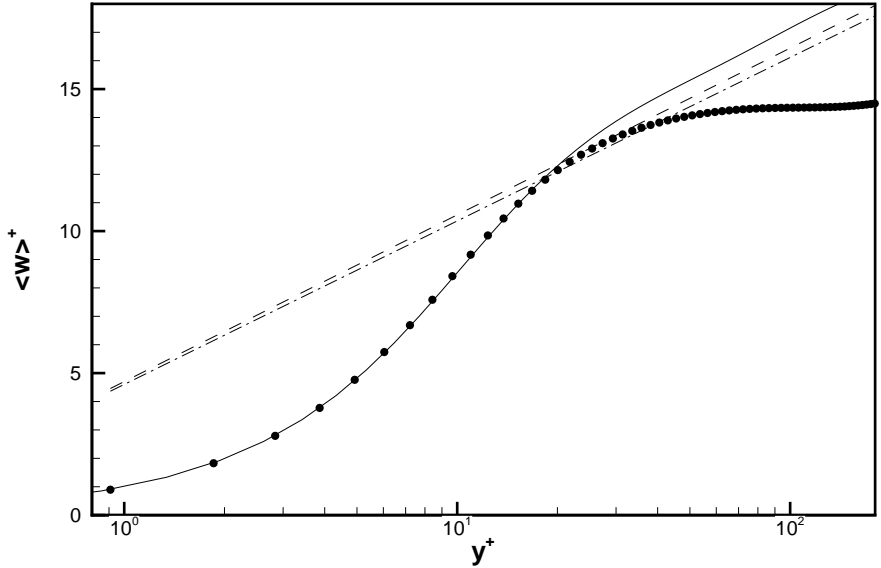}
\caption{Mean azimuthal velocity profile $\averfull{w}^+$ (symbols) over the inner wall, compared with logarithmic laws proposed in literature. Dashed line: plane Couette flow at $Re_\tau=52$ by \cite{bech-etal-1995}. Dash-dotted line: plane Couette flow at $Re_\tau=82.2$ by \cite{komminaho-lundbladh-johansson-1996}. Continuous line: plane channel flow at $Re_\tau=180$ by \cite{moser-kim-mansour-1999}.}
\label{fig:lotw-inner}
\end{figure}

The very existence of a logarithmic layer, where the mean velocity profile is described by:
\[
\averfull{w}^+ = \frac{1}{\kappa} \ln y^+ + B,
\]
as well as its characterization through the numerical values of the von K\'arm\'an constant $\kappa$ and the intercept $B$, are matters scholars have not yet agreed upon \cite{patel-sotiropoulos-1997}. Figure \ref{fig:lotw-inner} shows the azimuthal mean velocity profile, plotted in law-of-the-wall form for the inner wall only. The computed profile is compared to a few laws proposed in the literature. In particular for a plane Couette flow at $Re_\tau = 82.2$ the values $1/\kappa = 2.55$ and $B = 4.7$ have been suggested \cite{bech-etal-1995}, whereas for the same flow at $Re_\tau=52$ the values $1/\kappa = 2.5$ and $B=4.6$ have been used \cite{komminaho-lundbladh-johansson-1996}. The values for the plane channel flow at $Re_\tau=180$, after Kim, Moin \& Moser \cite{kim-moin-moser-1987}, are $1/\kappa = 2.5$ and $B=5.5$.

Figure \ref{fig:lotw-inner} reveals a rather limited extent of the logarithmic region, the relatively high value of $Re$ notwithstanding. (A rough estimate is that the present profile has constant slope for $20 \lesssim y^+ \lesssim 40$. The outer edge of the logarithmic layer can also be estimated from figure \ref{fig:lotw} as the position where the curves relative to the two walls start to diverge. On the contrary, the profile of the plane channel flow has constant slope in a range of nearly 100 wall units.) The values of $1/\kappa$ and $B$ from the plane channel case do not provide a good fit, whereas the ones from the plane Couette flow appear more suited to the present velocity profile. The best fitting line seems to be that by Ref.  \cite{bech-etal-1995}. The main information that can be drawn is thus to reinforce the statement by Smith \& Townsend  \cite{smith-townsend-1982}: ``no significant region of logarithmic variation of velocity can exist'' for ``any flow of Reynolds number less than $20000$'' in a Taylor--Couette flow with the outer cylinder at rest. According to a semi-empirical law \cite{bilgen-boulos-1973}, which relates the outer-scale Reynolds number to $Re_\tau$, this corresponds to $Re_\tau \approx 340$. As long as such information is not available, the collected evidence suggests that, at least for these relatively low values of $Re$, $\kappa$ should take values equal or slightly reduced compared to the plane channel flow. We recall however that curvature for the considered geometry at $\zeta=0.882$ is rather weak.

The observed form of the mean velocity profile is relevant to the ongoing discussion about the derivation of a friction law for the turbulent Taylor--Couette flow. In recent years, Lathrop et al \cite{lathrop-fineberg-swinney-1992} derived an approximate friction law based on the assumption of a logarithmic velocity profile, whereas Panton \cite{panton-1992} started from the assumption that the core region has constant angular momentum. The present results, which show to a very good approximation the presence of the constant angular momentum region,  appear to lend more support to the latter assumption. However, tough wider logarithmic layers could develop at higher $Re$, our data do not rule out the possibility that different theories (see for example Ref. \cite{eckhardt-grossmann-lohse-2000} and \cite{dubrulle-hersant-2002}) may yield a better friction law.

\begin{figure}
\centering
\includegraphics[width=\columnwidth]{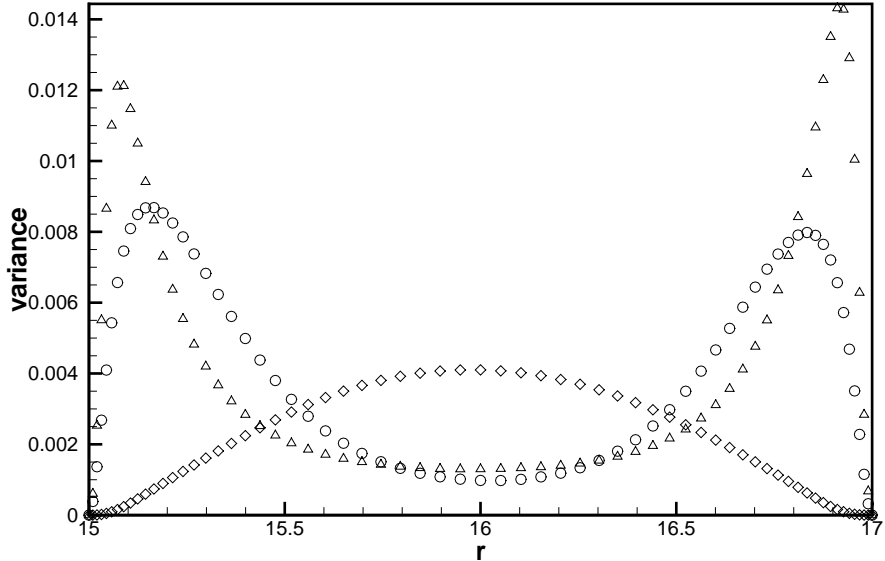}
\caption{Variance of velocity fluctuations across the gap. Triangles: streamwise component. Diamonds: wall-normal component. Circles: spanwise component. For better clarity only one every two points is shown.}
\label{fig:rms-outer}
\end{figure}
The variance of the velocity fluctuations is plotted in figure \ref{fig:rms-outer} across the whole gap. One observes the highest levels of fluctuations for the streamwise component, which is also the one with more marked asymmetry between the two walls and shows higher turbulence activity over the outer wall. The radial component peaks at the centerline with a symmetric profile, whereas the spanwise component is nearly symmetric and remarkably intense. It emerges clearly how the wall-normal component completely differs from the other ones in the central region of the gap.

\begin{figure}
\centering
\includegraphics[width=\columnwidth]{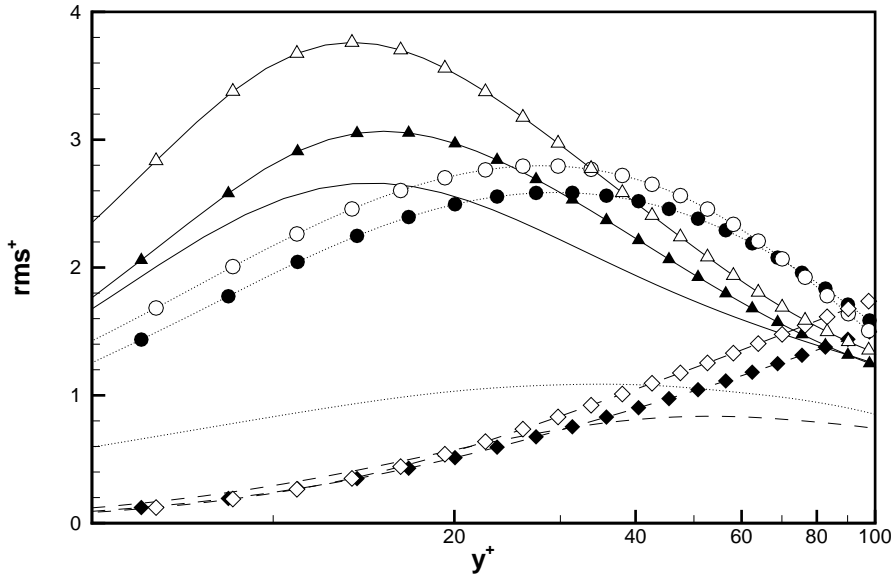}
\caption{Root-mean-square value of velocity fluctuations above both walls. Closed symbols: inner wall. Open symbols: outer wall. Lines without symbols: plane channel flow at $Re_\tau=180$ by \cite{kim-moin-moser-1987}. Triangles and continuous line: streamwise component. Diamond and dashed line: wall-normal component. Circles and dotted line: spanwise component. For clarity only one every two points is shown.}
\label{fig:rms}
\end{figure}
A different look can be obtained by plotting the root-mean-square values for both walls in local wall units against $y^+$ in figure \ref{fig:rms}, thus implicitly accounting for the different values of the friction Reynolds number. Convex and concave walls do still present discernible differences: in particular the turbulence activity as deduced by the level of streamwise velocity fluctuations is still larger over the outer wall. A similar but less pronounced behaviour is observed for the spanwise component. Inner scaling seems to work for the radial component, at least very near the wall: when the centerline is approached, the two profiles over the inner and outer wall appear to diverge.

We interpret the failure of friction velocity in scaling the r.m.s. profiles and the particular behaviour of the $v$ component as effects of the turbulent Taylor vortices. Near the wall they induce perturbations in the $u$ and $w$ components without affecting $v$, owing to the presence of the solid wall. In a sense, these perturbations behave similarly to the so-called inactive motions hypothesized by Townsend \cite{townsend-1976} to populate the turbulent boundary layer, and known to be responsible for the so-called anomalous scaling \cite{degraaf-eaton-2000,delalamo-etal-2004}. Figure \ref{fig:rms} reports also the r.m.s. value of velocity fluctuations for the plane channel flow, taken from \cite{moser-kim-mansour-1999}. The streamwise component is quite similar, while the wall-normal component is similar in the near-wall region only. The spanwise component in the channel flow presents a much lower level of activity, and is thus the component where the contribution by the turbulent Taylor vortices is the most relevant.

\begin{figure}
\centering
\includegraphics[width=\columnwidth]{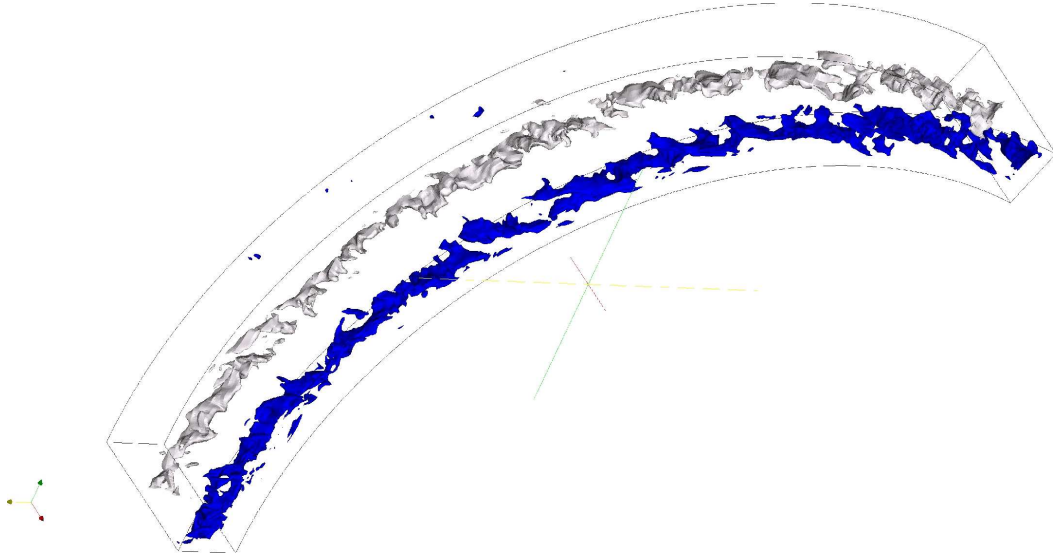}
\caption{Snapshot of an instantaneous flow field.  The bottom wall moves from left to right. Isosurfaces for the radial velocity component are shown, at levels of $\pm 0.12 W_i$: negative levels are in light gray, and positive in dark gray (blue).}
\label{fig:ttv}
\end{figure}
Figure \ref{fig:ttv} reports a snapshot of the whole flow field. A pair of turbulent Taylor vortices can be clearly observed. In laboratory experiments, they are usually visualized by resorting to passive tracers, whereas DNS makes all the flow variables easily available. Fig.\ref{fig:ttv} shows the entire computational domain: two isosurfaces where the radial velocity component attains a value of $\pm 0.12 W_i$ suggest the presence of two large-scale roll-like structures. It is evident how the vortices are strongly modulated by the noisy turbulent background, so that their boundary is somewhat blurred. They are however clearly identifiable, and extend down to the wall.

\section{Discussion and conclusions}
\label{sec:conclusions}

A direct numerical simulations of a Taylor--Couette flow in the fully turbulent regime has been presented. The value of the Reynolds number -- up to $Re_\tau \approx 190$ based on friction velocity, viscosity of the fluid and half gap width -- is high enough to compare turbulent statistics of the present flow and of planar pressure-driven flows.

Such a demanding simulation has required developing a numerical method designed {\em ad hoc}. It discretizes the governing equations with high accuracy and exploits the computing power of a parallel computing system, which is assembled from commodity hardware. The code, based on Fourier expansion in the homogeneous directions and on fourth-order, compact finite differences in the radial direction, has been validated by computing several physical quantities available from previous DNS or experiments in the laminar and transitional regimes, and has then made possible the first DNS of the Taylor--Couette flow in the fully turbulent regime.

A geometry with relatively small curvature has been considered. In comparing statistics to those from other wall turbulent flows, analogies and differences have been observed, the main source for the latter being the presence of the curvature-induced turbulent Taylor vortices. The mean velocity profile does not exhibit an equilibrium logarithmic region at the present value of $Re$, contrasting the pressure-driven flows. A core region with almost constant angular momentum has been revealed by our simulations, and the numerical value of this constant compares very well to experimental observations taken at different $Re$. The analysis of the root-mean-square values of velocity fluctuations, in comparison with the ones from a plane channel flow, reveals the lack of inner scaling, thus pointing to the existence of different physical processes contributing to turbulent fluctuations.

Two distinct sources of fluctuations can indeed be identified in the present flow: (i) the large-scale vortices, generated by an instability mechanism ultimately related to curvature, and (ii) the wall turbulence cycle, linked to the near-wall shear, which in the present context produces small-scale fluctuations and provides a noisy background for the vortices. A visualization of a snapshots of the flow field has shown how the large-scale structures are embedded in the flow and retain their basic shape, though blurred by the surrounding small-scale turbulence. The interaction of the small-scale turbulence with the large-scale structures, and the possibility of discerning and eventually separating their contributions to the flow statistics is an interesting long-term objective on which we are currently working.


\section*{Acknowledgments}
We acknowledge financial support by the Italian Ministry of University and Research through the PRIN project 2004 on {\em Turbulent flows over curved walls}.

\section*{Appendix}
\label{sec:details}

This Appendix provides some details concerning the entire procedure of extending the numerical method from what documented in Ref. \cite{quadrio-luchini-2002} to the one employed in the present simulations. The complete derivation can be found in Ref. \cite{pirro-2005}.

Since two directions are homogeneous, the first step consists in transforming the governing equations in Fourier space. Fourier-transformed quantities will be denoted with a hat; $\alpha$ and $m$ denote the axial and azimuthal wavenumbers, and $k^2=\alpha^2+m^2/r^2$. Note that $k$ depends on the radial coordinate. In \cite{quadrio-luchini-2002}, after manipulation of the Navier--Stokes equations in primitive variables the following equation for the Fourier-transformed radial vorticity $\heta$ was determined:
\begin{equation}
\frac{\partial \heta}{\partial t} =
\frac{1}{Re} \left( D D_*(\heta) - k^2 \heta
+ 2 \frac{i m}{r^2} D(\hu) + 2 \frac{m \alpha}{r^2} \hv \right) + 
\frac{im}{r} \hHU - i \alpha \hHW
\label{eq:eta}
\end{equation}

In Eq. (\ref{eq:eta}) the so-called Chandrasekhar notation is employed for radial derivatives:
\[
D(f) = \frac{\partial f}{\partial r}; \qquad D_*(f) = \frac{\partial
  f}{\partial r} + \frac{f}{r},
\]
and $\hHU$, $\hHW$ and (in the following Eq.(\ref{eq:v})) $\hHV$ denote the non-linear terms as written in the original primitive-variables equations.

After further manipulation, and leveraging the continuity equation, the temporal evolution of $\hv$ was shown in \cite{quadrio-luchini-2002} to be described by the following fourth-order differential equation:
\begin{multline}
  \frac{\partial}{\partial t} \left[\hv - D\left(\frac{1}{k^2}D_*(\hv)
    \right) \right] = \frac{1}{Re} D \left\{ \frac{1}{k^2} \left[
      k^2 D_*(\hv) -
\right. \right. \\ \left. \left. 
D_* D D_*(\hv) - 2 \frac{m^2}{r^3} \hv +
2 \frac{im}{r^2} D(\hw) - 2 \frac{im}{r^3} \hw \right] \right \} + \\
\frac{1}{Re} \left( -k^2 \hv + D D_*(\hv) - 2 \frac{im}{r^2}
    \hw \right) + \\
  D \left[ \frac{1}{k^2} \left( i \alpha\ \hHU + \frac{im}{r}
      \hHW \right) \right] + \hHV .
\label{eq:v}
\end{multline}

The entire differential system requires 6 boundary conditions; the no-slip and no-penetration conditions imposed at both walls imply $\hv=0$, $D(\hv)=0$ and $\heta=0$ at $r=\sR_i$ and $r=\sR_o$ \cite{orszag-israeli-deville-1986}.


Unlike the cartesian case, it can be noted that $\hu$ and $\hw$ appear in Eqs. (\ref{eq:eta}) and (\ref{eq:v}) through some of the viscous terms too (note that such terms vanish in the planar limit $r \rightarrow \infty$), owing to the form of the Laplacian operator in cylindrical coordinates. Hence, when a partially-implicit procedure is used for temporal integration, these curvature-related viscous terms cannot be solved for implicitly. The solution adopted in \cite{quadrio-luchini-2002} was to solve for them explicitly, since this was shown to imply no stability limitations.

Though a FD discretization is more convenient than a spectral one in terms of parallel computing, one has to deal with its inferior spatial accuracy. The accuracy can be made comparable to that of spectral methods by using compact, high-accuracy FD schemes \cite{lele-1992}. A further advantage is that the computational cost of compact schemes, that are usually implicit and require solving a system to evaluate a derivative, can be further reduced for the Navier--Stokes equations: in the present context, they can be made explicit, as observed as early as in 1953 by Thomas \cite{thomas-1953}, when third derivatives are absent in the governing equations.
We thus begin by eliminating third derivatives in Eq. (\ref{eq:v}) via the continuity equation. The simple substitution:
\[
D_* (\hv) = - i \alpha \hu - \frac{im}{r} \hw .
\]
does the job, and of course causes a few viscous terms to be moved into the part of the evolution equation (\ref{eq:v}) that is to be integrated explicitly in time. Similarly to the previously mentioned terms, this does not limit temporal stability, and the choice of the time step size is still dictated by considerations of temporal accuracy.

Now, if $a(r)$ indicates a generic $r$-dependent coefficient and $f$ is the unknown function, terms in the form $a D(f)$ do not lend themselves to a straightforward use of compact explicit schemes as in \cite{luchini-quadrio-2006}. Such terms must be rewritten after repeated integrations by parts, i.e. with substitutions of the type:
\begin{equation}
\label{eq:integration-by-parts}
a D (f) = D (a f) - D (a) f ; \qquad a D_* (f) = D_* (a f) - D (a) f .
\end{equation}

Note that integrating by parts introduces additional known coefficients like $D(a)$, depending on the radial coordinate only (directly and/or via the wavenumbers): one of the simplest among them is $D(1/k^2)$.

The procedure of repeated integrations by parts needed to massage Eqns. (\ref{eq:eta}) and (\ref{eq:v}) into a form suitable for the application of the compact finite-difference operators is now described in some detail.

In Eqn. (\ref{eq:v}), the first term which is integrated by parts following (\ref{eq:integration-by-parts}) is the time derivative, which becomes:
\[
\frac{\partial}{\partial t} \left[ - D \left(\frac{1}{k^2}D_*(\hv)
  \right) \right] = \frac{\partial}{\partial t} \left[ - D D_*
  \left( \frac{1}{k^2} \hv \right) + D \left(
  D ( \frac{1}{k^2} ) \right) \hv \right] .
\]

In the right-hand-side of Eqn. (\ref{eq:v}), perhaps the most complicated term is:
\[
D \left[ \frac{1}{k^2} \left( - D_* D D_* \hv \right) \right] ,
\]
where first the continuity equation must be invoked to cancel the third derivative. Repeated integrations by parts then allow the $r$-dependent coefficients to remain only in the innermost positions. After some algebra, the result is:
\begin{multline*}
- D \left[ \frac{1}{k^2} \left( D_* D D_* \hv \right) \right] = -
D D_* D D_* \left( \frac{1}{k^2} \hv \right) + D \left[ \frac{1}{r}
  D D \left(\frac{1}{k^2} \hv \right) \right] + \\
 - 2 D D_* \left[ \frac{1}{r} D
      \left(\frac{1}{k^2} \right) \hv \right] + D \left[ D D D
      \left(\frac{1}{k^2} \right) \hv \right] - D \left[ \frac{1}{r^2} D
        \left( \frac{1}{k^2} \right) \hv \right] + \\
- 3 D D_* \left[ D \left(\frac{1}{k^2}\right) \left( i \alpha \hu +
    \frac{im}{r} \hw \right) \right].
\end{multline*}

Having used the continuity equation, the last term has appeared which cannot be solved for implicitly, and must thus join the viscous curvature terms in the explicit part.

The last term of Eqn. (\ref{eq:v}) which needs manipulation is:
\[
D \left[ \frac{1}{k^2} \left( 2 \frac{im}{r^2} D (\hw) \right)
\right] = 2 im \left\{ D D \left( \frac{\hw}{k^2 r^2} \right) - D
\left[ \frac{1}{r^2} D \left(\frac{1}{k^2} \right) \hw \right] +
D \left( \frac{2}{k^2 r^3} \hw \right) \right\} .
\]

The same sequence of integration by parts must be carried out for Eqn. (\ref{eq:eta}) for the radial vorticity, arriving at the following substitution:
\[
2 \frac{im}{r} D(\hu) = 2 im \left[ D \left( \frac{u}{r^2} \right) + 2
  \frac{u}{r^3} \right] .
\]

Lastly, the nonlinear terms too contain radial derivatives, and some terms therein must be integrated by parts.

This procedure leads to the final, rather long form of the equations for $\hv$ and $\heta$, which lends itself to a discretization in the radial direction with explicit compact finite difference schemes of fourth-order accuracy over a five point stencil. It is written here in full for completeness, without time discretization for notational simplicity:
\begin{multline}
\label{eq:final-v}
\frac{\partial}{\partial t} \left[ \hv - D D_* \left( \frac{1}{k^2}
    \hv \right) + D \left( \hv D ( \frac{1}{k^2} ) \right) \right] =
\\
\frac{1}{Re} \left\{ 2 D D_* (\hv) - D D_* D D_* \left( \frac{1}{k^2}
    \hv \right) + D \left[ \frac{1}{r} D D \left(\frac{1}{k^2}\right)
    \hv \right] - 2 D D_* \left[ \frac{1}{r}
    D \left(\frac{1}{k^2}\right) \hv \right] +
\right. \\ \left.
+ D \left[ D D D \left(\frac{1}{k^2}\right) \hv \right] - D \left[
\frac{1}{r^2} D_1 \left(\frac{1}{k^2}\right) \hv \right] - 3 D D_*
\left[ D \left(\frac{1}{k^2}\right) \left( i \alpha \hu + \frac{im}{r}
    \hw \right) \right] +
\right. \\ \left.
- 2 m^2 D \left( \frac{1}{k^2 r^3} \hv \right) + 2 i m D
\left[\frac{1}{k^2 r^3} \hw \right] + 2 im \left[
  D D \left(\frac{1}{k^2 r^2} \hw \right) - D \left[ \frac{1}{r^2}
  D \left(\frac{1}{k^2}\right) \hw \right] \right] +
\right. \\ \left.
- k^2 \hv - 2 \frac{im}{r^2} \hw \right\} - i \alpha \left[ D D_*
\left( \frac{1}{k^2} \widehat{uv} \right) - D \left( D \left(\frac{1}{k^2}
    \right) \widehat{uv} \right) \right] - im \left[ D D
\left(\frac{1}{rk^2} \widehat{vw} \right) + 
\right. \\ \left.
+ 3 D \left( \frac{1}{r^2 k^2} \widehat{vw} \right) - D \left(
\frac{1}{r} D \left(\frac{1}{k^2} \right) \widehat{vw} \right) \right] +
D \left[
\frac{1}{k^2} \left( \alpha^2 \hu^2 + 2 \frac{\alpha m}{r} \widehat{uw} +
  \frac{m^2}{r^2} \hw^2 \right) \right] + \\ - i \alpha \widehat{uv} - D
  (\hv^2) - \frac{im}{r} \widehat{vw} -
\frac{1}{r} \hv^2 + \frac{1}{r} \hw^2 ;
\end{multline}

\begin{multline}
\label{eq:final-eta}
\frac{\partial \heta}{\partial t} =
\frac{1}{Re} \left\{ D D_*(\heta) - k^2 \heta + 2 i m \left[
    D \left(\frac{1}{r^2} \hu \right) + 2 \frac{\hu}{r^3} -
    \frac{i\alpha}{r^2} \hv \right] \right\} - i m \left[ \frac{i
    \alpha}{r} \hu^2 + 
\right. \\ \left.
D \left( \frac{\widehat{uv}}{r} \right) + \frac{2}{r^2} \widehat{uv} +
\frac{im}{r^2} \widehat{uw} \right] + i \alpha \left[ i \alpha \widehat{uw} +
D \left( \widehat{vw} \right) + \frac{im}{r} \hw^2 + \frac{2}{r}
\widehat{vw} \right].
\end{multline}

At this stage, the strategy for efficient parallel computing that has been successfully used in our cartesian Navier--Stokes solver, recently described as the Pipelined Linear System (PLS) method by Luchini and Quadrio in \cite{luchini-quadrio-2006}, can be implemented quite easily. The resulting code can run without modifications on the computing machine designed and built for our cartesian code, with obvious advantages.



\clearpage

\begin{table}
\centering
\begin{tabular}{ccccccccc}
                         & $L_x$  & $L_\theta$ & $N_x$ & $N_r$ & $N_\theta$ & $Re$   \\
``straight-vortex'' case & $4 h$  & $  \pi h $ &  64   &  64   &   32       & 60-190 \\
``wavy-vortex'' case     & $4 h$  & $2 \pi h$  &  64   &  64   &   32       & 255    \\
\end{tabular}
\caption{Size of the computational domain and grid resolution for the preliminary simulations in the ``straight-vortex'' and ``wavy-vortex'' regimes (see text).}
\label{tab:parameters-validation}
\end{table}

\begin{table}
\centering
\begin{tabular}{ccc}
$Re$ & here & $\zeta=0.5$ \cite{fasel-booz-1984} \\
60.0	& 16.7551	& 16.7551 \\
68.0	& 16.7551	& 16.7551 \\
70.0    & 17.1542	& 17.1537 \\
75.0	& 18.1634	& 18.1627 \\
80.0	& 19.0536	& 19.0527 \\
\end{tabular}
\caption{Comparison between the value of the torque $\sT$ computed by the present code and Ref. \cite{fasel-booz-1984} for $\zeta=0.5$.}
\label{tab:torque}
\end{table}

\begin{table}
\centering
\begin{tabular}{c|cccccccccc}
$S$	& 1	& 1.97	& 2.8	 & 3.7	& 4.6 	& 5.5	& 6.1	& 7	& 7.8 	& 8.5 \\
\hline
$N$ 	& 1	& 2	& 3	 & 4	& 5 	& 6	& 7	& 8	& 9 	& 10 \\
\end{tabular}
\caption{Parallel speedup $S$ of the code versus the number $N$ of the computing nodes. }
\label{tab:speedup}
\end{table}

\begin{table}
\centering
\begin{tabular}{cccccccccc}
Position    &$L_z $ & $L_\theta$&$L^+_z$& $L^+_{\theta}$ & $\Delta z^+$ & $y^+$ &
$(r \Delta \theta)^+$ \\ \\
inner wall  & $5 h$ & $9.37 \pi h$  & 946 & 5576 & 5.6 & 0.9 & 10.9 \\
centreline  & $5 h$ & $10 \pi h$    & 889 & 5588 & 5.2 & 4.6 & 10.9 \\
outer wall  & $5 h$ & $10.62 \pi h$ & 839 & 5600 & 4.9 & 0.8 & 10.9 \\
\end{tabular}
\caption{Size of the computational domain and grid resolution for the turbulent Taylor--Couette simulation. Wall units are computed on the basis of the friction velocity relative to each wall, as defined in \S\ref{sec:results}.}
\label{tab:parameters-turbulent}
\end{table}

\end{document}